\documentclass[aps,preprintnumbers,nofootinbib,superscriptaddress,11pt]{revtex4}
\usepackage[hypertex]{hyperref}
\usepackage[dvipdfmx]{graphicx} 
\usepackage{amsmath,amssymb,amsfonts,bm,cancel} 

\def\a{\alpha}    \def\d{\delta}     \def\th{\theta}      \def\n{\nu}               

\def\dg{\dagger}  \def\nn{\nonumber}

\newcommand{\vev}[1]{ \langle {#1} \rangle }
\def\abs#1{\left| #1\right|}

\newcommand{\row}[2]{ \begin{pmatrix}  #1 & #2   \end{pmatrix}  }
\newcommand{\column}[2]{ \begin{pmatrix}  #1 \\ #2 \\  \end{pmatrix} }

\newcommand{\diag}[2]{ \begin{pmatrix}  #1 & 0 \\ 0 & #2 \\   \end{pmatrix}  }

\newcommand{\Diag}[3]{ \begin{pmatrix} #1 & 0 & 0 \\ 0 & #2 & 0 \\ 0 & 0 & #3 \\\end{pmatrix}}

\begin{document}


\title{\large General solutions for TM$_{1,2}$ mixing in the minimal type-I seesaw mechanism \\ and phenomenological constraints to Yukawa matrix of neutrinos}

\preprint{STUPP-23-265}
\author{Masaki J. S. Yang}
\email{mjsyang@mail.saitama-u.ac.jp}
\affiliation{Department of Physics, Saitama University, 
Shimo-okubo, Sakura-ku, Saitama, 338-8570, Japan}



\begin{abstract} 

In this paper, we concisely represent the general solution for the TM$_{1,2}$ mixing in the type-I minimal seesaw mechanism    and show phenomenological constraints of the Yukawa matrix of neutrinos $Y$. 
First, conditions of $Z_{2}$ symmetry associated with TM$_{1,2}$ restrict two complex parameters.  
In addition, three complex parameters are constrained from five phenomenological parameters (two neutrino masses $m_{i}$, the mixing angle $\theta_{13}$, the Dirac phase $\delta$, and the Majorana phase $\alpha$) and the overall phase.
As a result, there is only one free complex parameter for $Y$, except for the right-handed neutrino masses $M_{1,2}$.

For the TM$_{1}$ mixing with normal hierarchy, the symmetry condition is imposed only on $Y$, and the physical parameters depend on $M_{R}$. In other situations, the dimensionless parameters $(\theta_{13}, \delta$ and $\alpha)$ do not depend on $M_{R}$ but only on certain combinations of the components of $Y$.
 
\end{abstract} 

\maketitle

\section{Introduction}

Discrete symmetry has been implemented in a number of studies for the neutrino mass  \cite{Ishimori:2010au}. 
This is because its characteristic mixing pattern is close to the tri-bi-maximal mixing (TBM) \cite{Harrison:2002er}, which involves $Z_{2} \times Z_{2}$ symmetry \cite{Lam:2006wm,Lam:2007qc,Lam:2008rs}.
Since the finite $\th_{13}$ was observed \cite{DayaBay:2012fng}, the TM$_{1,2}$ mixing is still being discussed today \cite{Albright:2008rp,Albright:2010ap,He:2011gb,Luhn:2013lkn,Li:2013jya,King:2019vhv,Krishnan:2020xeq}. 
Here one of the two $Z_{2}$ symmetries associated with the TBM is remaining unbroken. 
In particular, the $Z_{2}$ symmetry associated with TM$_{2}$ (the trimaximal mixing \cite{Harrison:2002kp,Friedberg:2006it,Bjorken:2005rm, He:2006qd,Grimus:2008tt,Channey:2018cfj,Bao:2021zwu,Yang:2021xob}) is called the {\it magic} symmetry,
and the matrix $m$ is called {\it magic} in which the row sums and the column sums are all equal to a number $\a$ \cite{Lam:2006wy}. 

These TM$_{1,2}$ mixing has also been investigated \cite{Shimizu:2017fgu, Shimizu:2017vwi, Zhao:2021efc} in models with the type-I seesaw mechanism \cite{Minkowski:1977sc,GellMann:1980v,Yanagida:1979as, Mohapatra:1979ia}. 
However, the discussion has been limited to a certain basis and/or specific structures. 
In a previous paper \cite{Yang:2022yqw}, 
we considered the conditions of TM$_{1,2}$ mixing  in a general basis of $M_{R}$ for the minimal type-I seesaw model 
\cite{Ma:1998zg, King:1998jw, Frampton:2002qc, Xing:2020ald, Guo:2003cc,Barger:2003gt, Mei:2003gn, Chang:2004wy, Guo:2006qa, Kitabayashi:2007bs, He:2009pt, Yang:2011fh, Harigaya:2012bw, Kitabayashi:2016zec, Bambhaniya:2016rbb, Li:2017zmk, Liu:2017frs, Nath:2018hjx, Barreiros:2018bju, Nath:2018xih,  Wang:2019ovr, Zhao:2020bzx}. 
Since the conditions in the previous paper were rather complicated, 
in this paper, we organize the conditions for the TM$_{1,2}$ mixing into a more concise form 
and determine all the constraints that come from phenomenological parameters.
 
This paper is organized as follows. 
The next section gives a review of TM$_{1,2}$ mixing and its conditions in the minimal type-I seesaw mechanism. 
In Sec.~3, we analyze phenomenological constraints of  TM$_{1,2}$ mixing. 
The final section is devoted to a summary.

\section{Minimal type-I seesaw mechanism with TM$_{1,2}$ mixing and its conditions}

In this section, we organize conditions for the type-I minimal seesaw model \cite{Ma:1998zg, King:1998jw, Frampton:2002qc} to have the TM$_{1,2}$ mixing \cite{Albright:2008rp,Albright:2010ap,He:2011gb} defined as 
\begin{align}
U_{\rm TM1} = U_{\rm TBM} U_{23} \, , ~~~
U_{\rm TM2} = U_{\rm TBM} U_{13} \, . ~~~
\end{align}
Here,  
\begin{align}
U_{\rm TBM} 
= 
\begin{pmatrix}
 \sqrt{\frac{2}{3}} & \frac{1}{\sqrt{3}} & 0 \\
 -\frac{1}{\sqrt{6}} &\frac{1}{\sqrt{3}} & \frac{1}{\sqrt{2}} \\
 -\frac{1}{\sqrt{6}} & \frac{1}{\sqrt{3}} & - \frac{1}{\sqrt{2}} \\
\end{pmatrix} , 
~~~
U_{23} = 
\begin{pmatrix}
1 & 0 & 0 \\
0 & c_{\th} & s_{\th} e^{- i\phi} \\
0 & - s_{\th} e^{i \phi} & c_{\th}
\end{pmatrix} ,
~~~ 
U_{13} =
\begin{pmatrix}
c_{\th} & 0 & s_{\th} e^{- i \phi} \\
0 & 1 & 0 \\
- s_{\th} e^{i \phi} & 0 & c_{\th}
\end{pmatrix} \, , 
\label{U123}
\end{align}
with $c_{\th} \equiv \cos \th , \, s_{\th} \equiv \sin \th$. 
By using the notation in the PDG parameterization $c_{ij} \equiv \cos \th_{ij} \, , s_{ij} \equiv \sin \th_{ij}$, the value of $s_{13}$ is
\begin{align}
s_{13}^{2} = \sin^{2} \th^{\rm TM_{1}} / 3  =  2 \sin^{2} \th^{\rm TM_{2}} /  3 \, . 
\end{align}
The Jarlskog invariant \cite{Jarlskog:1985ht} is evaluated as \cite{Bjorken:2005rm, Xing:2006ms, He:2006qd, Albright:2008rp, Albright:2010ap}
\begin{align}
J^{\rm TM_{1}} = { c_{\th} s_{\th} \sin \phi  \over 3 \sqrt{6}} \, , ~~~ 
J^{\rm TM_{2}} = {c_{\th} s_{\th} \sin \phi \over 3 \sqrt 3} \, . \label{Jarlskog}
\end{align}
Since the invariant with the Dirac phase $\d$ is written as $J = c_{12} s_{12} c_{23} s_{23} c_{13}^{2} s_{13} \sin \d$,  $\sin \phi \simeq \sin \d$ holds in both cases \cite{Bjorken:2005rm}. 
Furthermore, $\cos \d$ is also written by rephasing invariants of MNS matrix $U$; 
\begin{align}
\cos \d & =  { |U_{22}|^{2} - (s_{12} s_{13} s_{23})^{2} - (c_{12} c_{23})^{2} \over - 2 s_{12} s_{13} s_{23} c_{12} c_{23} }    \\
&= {|U_{22}|^{2} (1 - |U_{13}|^{2})^{2} - |U_{13}|^{2} |U_{12}|^{2} |U_{23}|^{2} - |U_{11}|^{2} |U_{33}|^{2} \over - 2 |U_{13}| |U_{12}| |U_{23}| |U_{11}| |U_{33}|} \, . 
\end{align}
Specific substitution yields 
$\cos \d \simeq {\rm sign}(s_{\th}^{\rm TM_{1}}) \cos \phi^{\rm TM_{1}} \simeq \cos \phi^{\rm TM_{2}} $, and relations between CP phases are
\begin{align}
\d \simeq \phi^{\rm TM_{1}} + \arg ({\rm sign}(s_{\th}^{\rm TM_{1}})) \simeq \phi^{\rm TM_{2}} \, .  
\label{phases}
\end{align}

It is well known that these TM$_{1,2}$ mixings accompany the $Z_{2}$ symmetry for the mass matrix $m_{\n}$ \cite{Lam:2006wm,Lam:2007qc,Lam:2008rs}. 
Let $\bm v_{i}$ be the column vectors of the TBM matrix, 
$(\bm v_{1} \, , \bm v_{2} \, , \bm v_{3}) \equiv U_{\rm TBM}$.
The TM$_{1,2}$ mixing is automatically predicated when the mass matrix of neutrinos $m_{\n}$ has a $Z_{2}$ symmetry generated by $S_{1,2} \equiv 1 - 2 \bm v_{1,2} \otimes \bm v_{1,2}^{T} $,
\begin{align}
S_{1,2} \, m_{\n} \, S_{1,2} = m_{\n} \, , ~~ 
S_{1} = 
{1\over 3}
\begin{pmatrix}
-1 & 2 & 2 \\
2 & 2 & -1 \\
2 & -1 & 2 \\
\end{pmatrix} \, ,
~~~
S_{2} = 
{1\over 3}
\begin{pmatrix}
1 & -2 & -2 \\
-2 & 1 & -2 \\
-2 & -2 & 1 \\
\end{pmatrix} \, . 
\label{S12sym} 
\end{align}
These symmetries fix certain eigenvectors of $m_{\n}$ to $\bm v_{1,2}$. 
These generators are diagonalized in the TBM basis and decompose $m_{\n}$ into two eigenspaces with charge $\pm 1$.
\begin{align}
U_{\rm TBM}^{\dg} S_{1} U_{\rm TBM}^{*} = \Diag{-1}{1}{1} \, , ~~ 
U_{\rm TBM}^{\dg} S_{2} U_{\rm TBM}^{*} = \Diag{1}{-1}{1} \, . 
\label{diagS12}
\end{align}

First, let us consider conditions under which the minimal type-I seesaw mechanism has TM$_{1,2}$ mixing 
and these $Z_{2}$ symmetries.
For simplicity, the vacuum expectation value (vev) of the Higgs field is set to unity. 
The Yukawa matrix $Y$ in the minimal seesaw model and the Majorana mass matrix of the right-handed neutrinos $\n_{R i}$ are given by 
\begin{align}
Y = 
\begin{pmatrix}
a_{1} & b_{1} \\
a_{2} & b_{2} \\
a_{3} & b_{3} \\
\end{pmatrix} ,
%
~~~ 
M_{R} =
\begin{pmatrix}
M_{11} & M_{12} \\ M_{12} & M_{22}
\end{pmatrix} . 
%
\label{1}
\end{align}
The Yukawa matrix in the TBM basis is also defined as
\begin{align}
Y_{\rm TBM} \equiv 
U_{\rm TBM}^{\dg} Y \equiv
\begin{pmatrix}
A_{1} & B_{1} \\
A_{2} & B_{2} \\
A_{3} & B_{3} \\
\end{pmatrix} 
\equiv (\bm A \, , \bm B) \, .
\end{align}
By the minimal type-I seesaw mechanism, 
 the mass matrix of light neutrinos $m_{\rm TBM}$ (in the TBM basis) is 
\begin{align}
m_{\rm TBM} & = 
U^{\dg}_{\rm TBM} m_{\n} U^{*}_{\rm TBM}  = 
Y_{\rm TBM} M_{R}^{-1} Y_{\rm TBM}^{T} \label{8}    \\
&= {1 \over \det M_{R}} 
\bigg( \bm A \otimes (M_{22} \bm A^{T} - M_{12} \bm B^{T} )
-  \bm B \otimes (M_{12} \bm A^{T} - M_{11} \bm B^{T}) \bigg ) \, .  \label{mnTBM}
\end{align}
The symmetries~(\ref{diagS12})  by $S_{1, 2}$ require zero textures $(m_{\rm TBM})_{12} = (m_{\rm TBM})_{13} = 0$ or $(m_{\rm TBM})_{21} = (m_{\rm TBM})_{23} = 0$ \cite{Shimizu:2017fgu}. 
\begin{align}
{\rm TM}_{1}&: 
\row{A_{1}}{B_{1}}
\begin{pmatrix}
M_{22} & - M_{12} \\ - M_{12} & M_{11}
\end{pmatrix}
\column{A_{2,3}}{B_{2,3}} = 0 \, ,  \label{cond1} \\
{\rm TM}_{2}&: 
\row{A_{2}}{B_{2}}
\begin{pmatrix}
M_{22} & - M_{12} \\ - M_{12} & M_{11}
\end{pmatrix}
\column{A_{1,3}}{B_{1,3}} = 0 \, . \label{cond2}
\end{align}
Since they are orthogonality conditions for two 2-dimensional vectors (without Hermitian conjugation),  all solutions can be classified into two types \cite{Yang:2022yqw}. 
\begin{enumerate}
\item Chiral solution: 

\begin{enumerate}
\item[(1)] TM$_{1}$: $S_{1} Y = \pm Y , ~~ A_{1} = B_{1} = 0$ or $A_{2,3} = B_{2,3} = 0\,$. 
\item[(2)] TM$_{2}$: $S_{2} Y = \pm Y , ~~ A_{2} = B_{2} = 0$ or $A_{1,3} = B_{1,3} = 0\,$. 
\end{enumerate}
This chiral symmetry is associated with the massless mode  \cite{Petcov:1984nz,  Adhikari:2010yt, Yang:2023ixv}.
For the singular values of light neutrinos $m_{i}$,  
$S_{1} Y = + Y$ with $m_{1} = 0$ leads to the normal hierarchy (NH). 
The case $S_{1} Y = - Y $ with $m_{2,3} = 0$ is phenomenologically excluded. 
For the TM$_{2}$,  the Yukawa matrix $Y$ cannot have the chiral symmetry by $S_{2}$ because it predicts $m_{2} = 0$ or $m_{1,3} = 0$. 
Therefore, this chiral solution is limited to only the TM$_{1}$ with NH.

\item Non-trivial solution\footnote{In Ref.~\cite{Yang:2022yqw}, the solution for $B_{1} = 0$ or $B_{2,3} = 0$ is classified in another class that is merged to this class. }:

If all two-dimensional vectors $(A_{i} \, , B_{i})$ are not the zero vector $\bm 0$,  
the orthogonality condition leads to 
\begin{enumerate}
\item[(1)] TM$_{1}$:  
\begin{align}
{\rm TM}_{1}&: 
\column{A_{2}}{B_{2}} \propto \column{A_{3}}{B_{3}} \propto 
\begin{pmatrix}
M_{11} & M_{12} \\ M_{12} & M_{22}
\end{pmatrix}
\column{- B_{1}}{A_{1}}  ,  \label{condTM1}
\end{align}
This solution requires $m_{1} \neq 0$ and automatically leads to IH with $m_{3} = 0$.
Explicitly, this condition can be written as\footnote{If the denominator diverges, we just consider the reciprocal.}
\begin{align}
{A_{2} \over B_{2}} = {A_{3} \over B_{3}} = {-M_{11} B_{1} + M_{12} A_{1} \over - M_{12} B_{1} + M_{22} A_{1}} \, , 
\end{align}
and it agrees with the results of \cite{Yang:2022yqw}. 
 
\item[(2)] TM$_{2}$:  
\begin{align}
{\rm TM}_{2}&:  
\column{A_{1}}{B_{1}} \propto \column{A_{3}}{B_{3}} \propto 
\begin{pmatrix}
M_{11} & M_{12} \\ M_{12} & M_{22}
\end{pmatrix}
\column{- B_{2}}{A_{2}}  .  \label{condTM2}
\end{align}
TM$_{2}$ has two solutions with NH and IH.
\end{enumerate}

\end{enumerate}
%

\section{Phenomenological constraints on the Yukawa matrix}

In this section, we investigate how the Yukawa matrix of neutrinos $Y$ with the above conditions of  TM$_{1,2}$ is constrained from phenomenology.
There is five phenomenological parameters; the light neutrino masses $m_{1 \, \rm or \, 3 } , m_{2}$, the mixing angle $\th$ of TM$_{1,2}$, the Dirac phase $\d$ and the Majorana phase $\a$. 
If we add the unphysical overall phase, the three complex parameters are restricted from observables. 
Therefore, except for the masses of right-handed neutrinos $M_{1,2}$, 
the degree of freedom in the Yukawa matrix is only one complex parameter.

\subsection{Chiral solution of TM$_{1}$}

This solution exists only in TM$_{1}$ and the structure of the Yukawa matrix in TBM basis is determined by the solution $S_{1} Y = Y\,, A_{1} = B_{1} = 0$;
\begin{align}
Y_{0} = 
\begin{pmatrix}
0 & 0 \\
A_{2} & B_{2} \\
A_{3} & B_{3}
\end{pmatrix} \, . 
\end{align}
Although values of $A_{2,3}$ and $B_{2,3}$ depend on the basis of $M_{R}$, 
these zero textures associated with the chiral symmetry are preserved.
Therefore, suppose $M_{R}$ is in the diagonal basis, 
\begin{align}
m_{\rm TBM}^{(0)} = Y_{0} M_{R}^{-1} Y_{0}^{T} 
= 
\begin{pmatrix}
0 & 0 & 0  \\
0 & {A_{2}^{2} \over M_{1}} + {B_{2}^{2} \over M_{2}} & {A_{2} A_{3} \over M_{1}} + {B_{2} B_{3} \over M_{2}}  \\
0 & {A_{2} A_{3} \over M_{1}} + {B_{2} B_{3} \over M_{2}}  & {A_{3}^{2} \over M_{1}} + {B_{3}^{2} \over M_{2}}   \\
\end{pmatrix} \, . 
\label{mTBMchiral}
\end{align}

Only this solution requires solving a complicated quadratic equation to find the exact singular value and mixings such as in Ref.~\cite{Fujihara:2005pv}.
However, this solution is expressed by a complex orthogonal matrix $O$ of the Casas--Ibarra representation \cite{Casas:2001sr}. 
Henceforth, the computation is restricted to the 2-3 submatrix. 
By defining $\tilde m_{2} \equiv m_{2} e^{i \a}$ with the only Majorana phase $\a$ associated with $m_{2}$, 
\begin{align}
\begin{pmatrix}
A_{2}  & B_{2}  \\
A_{3}  & B_{3}  \\
\end{pmatrix}
= 
\begin{pmatrix}
 c_{\th} & s_{\th} e^{- i\phi} \\
 - s_{\th} e^{ i \phi}  & c_{\th}
\end{pmatrix}
\diag{\sqrt {\tilde m_{2}} }{\sqrt {m_{3}} }
\begin{pmatrix}
c_{z} & -s_{z} \\
\pm s_{z} & \pm c_{z} \\
\end{pmatrix} 
\diag{\sqrt{M_{1}}}{\sqrt{M_{2}}} ,
\end{align}
where $z$ is an arbitrary complex parameter and $\pm$ denotes two independent solutions.

The phenomenological constraint of $Y_{0}$ is identified by matrix operations. 
When the orthogonal matrix $O$ is represented by
\begin{align}
O = 
\begin{pmatrix}
c_{z} & -s_{z} \\
\pm s_{z} & \pm c_{z} \\
\end{pmatrix}
=
\diag{\sqrt {\tilde m_{2}^{-1}}}{\sqrt {m_{3}^{-1}}} U_{23}^{\dg} \row{ \bm A \over \sqrt{M_{1}} }{\bm B \over \sqrt{M_{2}}}
 ,  
\end{align}
the following non-trivial relationship exists between the first and second columns of $O$,
\begin{align}
\diag{\sqrt {\tilde m_{2}^{-1}}}{\sqrt {m_{3}^{-1}}} U_{23}^{\dg}  {\bm A \over \sqrt{M_{1}}}
& = \column{c_{z}} {\pm s_{z}}  
= 
\begin{pmatrix}
0 & \pm 1 \\
\mp 1 & 0 
\end{pmatrix}
\column{- s_{z}} {\pm c_{z}}  \nn \\
&= \pm
\begin{pmatrix}
0 &  1 \\
- 1 & 0 
\end{pmatrix}
\diag{\sqrt {\tilde m_{2}^{-1}}}{\sqrt {m_{3}^{-1}}} U_{23}^{\dg}  {\bm B \over \sqrt{M_{2}} } \, .   
\label{cnstr}
\end{align}
Thus, by defining $Y_{0}'$ in the diagonalized basis of $m_{\n}$, 
\begin{align}
Y_{0}'  \equiv  U_{23}^{\dg} Y_{0}
%
\equiv 
\begin{pmatrix}
A_{2}' & B_{2}' \\
A_{3}' & B_{3}' 
\end{pmatrix}  , 
\end{align}
Eq.~(\ref{cnstr}) yields 
%
%
%
\begin{align}
\column{A_{2}'}{A_{3}'} = 
\pm \sqrt{M_{1} \over M_{2}} 
\begin{pmatrix}
0 &  \sqrt{\tilde m_{2} \over  m_{3}} \\
- \sqrt{m_{3} \over \tilde m_{2}} & 0 
\end{pmatrix}
\column{B_{2}'}{B_{3}'} 
 = 
\pm \sqrt{M_{1} \over M_{2} \tilde m_{2} m_{3}} 
\column{\tilde m_{2} B_{3}'}{ - m_{3}B_{2}'} .
\end{align}
From $s_{z}^{2} + c_{z}^{2} = 1$, the following constraint exists between $B_{2,3}'$
\begin{align}
{(B_{2}')^{2} \over \tilde m_{2}} + {(B_{3}')^{2} \over m_{3}} = M_{2} \, . 
\end{align}
Eventually, the chiral solution $Y_{0}$ for  $\det O = \pm1$ in the TBM basis
is
\begin{align}
%
%
%
Y_{0} = 
\begin{pmatrix}
A_{2} & B_{2} \\
A_{3} & B_{3} 
\end{pmatrix}
& = U_{23} Y_{0}'
= 
\begin{pmatrix}
 c_{\th} & s_{\th} e^{- i\phi} \\
 - s_{\th} e^{ i \phi}  & c_{\th}
\end{pmatrix}
\begin{pmatrix}
\pm \sqrt{M_{1} \tilde m_{2} \over M_{2} m_{3}} B_{3}' & B_{2}' \\
\mp \sqrt{M_{1} m_{3} \over M_{2} \tilde m_{2} } B_{2}' & B_{3}' 
\end{pmatrix} .
\label{25}
\end{align}
Three complex parameters of the Yukawa matrix are restricted. 

\subsection{Nontrivial solutions of TM$_{1}$}

The nontrivial solution of TM$_{1}$ is IH with $m_{1} \neq 0$. 
From the condition~(\ref{condTM1}), the form of $Y$ is  restricted to
\begin{align}
Y_{1} = 
\begin{pmatrix}
A_{1} & B_{1} \\
0 & 0 \\
0 & 0 \\
\end{pmatrix}  
+
%
\begin{pmatrix}
0 & 0 \\
- c_{2} B_{1} & c_{2} A_{1} \\
- c_{3} B_{1} & c_{3} A_{1} \\
\end{pmatrix} M_{R}  \, ,
\label{YTM1}
\end{align}
where $c_{i}$ are arbitrary complex coefficients.
Although the values of $A_{1}, B_{1}$ are changed by a basis transformation, 
the texture is independent on  the basis of $M_{R}$. 
For a two-dimensional unitary matrix $U_{2}$,
\begin{align}
Y_{1}' = Y_{1} U_{2} &=
\begin{pmatrix}
A_{1} & B_{1} \\
0 & 0 \\
0 & 0 \\
\end{pmatrix}  
U_{2}
+
%
\begin{pmatrix}
0 & 0 \\
- c_{2} B_{1} & c_{2} A_{1} \\
- c_{3} B_{1} & c_{3} A_{1} \\
\end{pmatrix} U_{2}^{*} U_{2}^{T} M_{R} U_{2} \\
& = 
\begin{pmatrix}
A_{1}' & B_{1}' \\
0 & 0 \\
0 & 0 \\
\end{pmatrix}  
+
%
\begin{pmatrix}
0 & 0 \\
- c_{2} B_{1}' & c_{2} A_{1}' \\
- c_{3} B_{1}' & c_{3} A_{1}' \\
\end{pmatrix}  M_{R}' \, .
\label{Utrf}
\end{align}
This is because the symmetry conditions~(\ref{cond1}) and (\ref{condTM1}) are covariants under basis transformation.
Moreover, the sign degree of freedom of $\det O$ can be absorbed into $c_{2,3}$.

As in the chiral solution, $M_{R}$ is taken to be diagonal.
Since the product of the two terms in $Y_{1}$ obviously orthogonal in the seesaw mechanism,  
the mass matrix $m_{\rm TBM}$ in the TBM basis is 
\begin{align}
m_{\rm TBM}^{(1)} &= 
Y_{1} M_{R}^{-1} Y_{1}^{T} \\
& = 
\Diag{ {A_{1}^{2} \over M_{1}} + {B_{1}^{2} \over M_{2}} }{0}{0}
+ (M_{2} A_{1}^{2} + M_{1}  B_{1}^{2}) 
\begin{pmatrix}
0 & 0 & 0 \\
0& c_{2}^{2} & c_{2} c_{3} \\
0 & c_{2} c_{3} & c_{3}^{2} \\
\end{pmatrix} \, . 
\label{mTM1}
\end{align}
The mass singular values are 
\begin{align}
m_{1} = \abs{{A_{1}^{2} \over M_{1}} + {B_{1}^{2} \over M_{2}} } \, , ~~
m_{2} = \abs{ M_{2} A_{1}^{2} + M_{1}  B_{1}^{2} } (|c_{2}|^{2} + |c_{3}|^{2}) \, , ~~ 
m_{3} = 0 \, . 
\label{miTM1}
\end{align}
Note that the 1-1 element is equal to $ {1\over M_{1}M_{2}} (M_{2} A_{1}^{2} + M_{1} B_{1}^{2})$. Thus, the overall phase of $ (M_{2} A_{1}^{2} + M_{1} B_{1}^{2})$ can be removed and does not affect the physical quantity. 
In particular, considering the division of singular values, we obtain a constraint on $ |c_{i}|^{2}$;  
\begin{align}
{m_{2} \over m_{1}} =  M_{1} M_{2} (|c_{2}|^{2} + |c_{3}|^{2})  \, , ~~
|c_{2}|^{2} + |c_{3}|^{2} = {m_{2} \over  m_{1} \det M_{R}} \, . 
\label{ratio12}
\end{align}

The unitary matrix that diagonalizes the mass matrix~(\ref{mTM1}) is
\begin{align}
 U^{\dg} = 
%
\begin{pmatrix}
1 & 0 & 0 \\
0 & {|c_{2}| \over \sqrt{|c_{2}|^{2} + |c_{3}|^{2}} } &  {|c_{3}| \over \sqrt{|c_{2}|^{2} + |c_{3}|^{2}} }\\
0 & - {|c_{3}| \over \sqrt{|c_{2}|^{2} + |c_{3}|^{2}} }& {|c_{2}| \over \sqrt{|c_{2}|^{2} + |c_{3}|^{2}}  }\\
\end{pmatrix}
\Diag{1}{e^{ - i \arg c_{2}}}{e^{ - i \arg c_{3}}} . 
\end{align}
By using a freedom of phase redefinition for $m_{3} = 0$ and 
comparing this to $U_{23}$~(\ref{U123}), 
\begin{align}
U_{23} P \equiv
\begin{pmatrix}
1 & 0 & 0 \\
0 & c_{\th} & s_{\th} e^{- i\phi} \\
0 & - s_{\th} e^{i \phi} & c_{\th}
\end{pmatrix} 
\Diag{1}{e^{i \a/2}}{1} 
=
\begin{pmatrix}
1 & 0 & 0 \\
0 & {|c_{2}| e^{i \arg c_{2} }  \over \sqrt{|c_{2}|^{2} + |c_{3}|^{2}} } & - {|c_{3}| e^{i (\arg c_{2} - \arg c_{3})}\over \sqrt{|c_{2}|^{2} + |c_{3}|^{2}} }\\
0 &  {|c_{3}| e^{i \arg c_{3} } \over \sqrt{|c_{2}|^{2} + |c_{3}|^{2}} }& {|c_{2}| \over \sqrt{|c_{2}|^{2} + |c_{3}|^{2}}  }\\
\end{pmatrix}  \, . 
\end{align}
Thus, the parameters of the matrix~(\ref{U123}) diagonalizing $m_{\rm TBM}^{(1)}$ and the Majorana phase are,
\begin{align}
\tan \th^{\rm TM_{1}} = - \abs{c_{3} \over c_{2}}  \simeq - \sqrt 3 \, s_{13} \, ,  ~~ \phi = \arg{c_{3} \over c_{2}} \simeq \d + \pi \,  \, ,  ~~ 
\a = 2 \arg c_{2} \, . 
\end{align}
To simplify the notation of $\phi$, we choose $\th < 0$. 
The $\d + \pi$ follows from Eq.~(\ref{phases}).  
In particular, these dimensionless parameters do not depend on $M_{R}$, but only on  ratios between the second and third rows of $Y_{1}$. 
By substituting these parameters to the Yukawa matrix~(\ref{YTM1}),
\begin{align}
Y_{1} = 
\begin{pmatrix}
A_{1} & B_{1} \\
0 & 0 \\
0 & 0 \\
\end{pmatrix}  
+
e^{i \a /2} \sqrt {m_{2} \over m_{1} \det M_{R} } 
\begin{pmatrix}
0 & 0 \\
- c_{\th}^{\rm TM_{1}} B_{1} & c_{\th}^{\rm TM_{1}}  A_{1} \\
s_{\th}^{\rm TM_{1}} e^{i \phi } B_{1} & - s_{\th}^{\rm TM_{1}} e^{i \phi} A_{1} \\
\end{pmatrix} M_{R} \, .
\end{align}
This notation is independent of the basis of $M_{R}$ and holds in a general basis.
Since $A_{1}$ and $B_{1}$ are constrained by $m_{i}~(\ref{miTM1})$ and the overall phase is unphysical, the only free parameter is a single complex number $(A_{1}$ or $B_{1})$. 
%
%
Furthermore, by substituting $s_{\th}^{\rm TM_{1}} e^{i \phi} \simeq - \sqrt{3} s_{13} e^{i(\d + \pi)}$, it can be displayed in  parameters of PDG representation.

\subsection{Nontrivial solution of TM$_{2}$}

Similarly, from the condition~(\ref{condTM2}), the form of the Yukawa matrix with TM$_{2}$ is restricted to 
\begin{align}
Y_{2} = 
\begin{pmatrix}
0 & 0 \\
A_{2} & B_{2} \\
0 & 0 \\
\end{pmatrix} 
+  
\begin{pmatrix}
- c_{1} B_{2} & c_{1} A_{2} \\
0 & 0 \\
- c_{3} B_{2} & c_{3} A_{2} \\
\end{pmatrix} M_{R} \, . 
\end{align}
The mass matrix in the TBM basis is
\begin{align}
m_{\rm TBM}^{(2)} &=  Y_{2} M_{R}^{-1} Y_{2}^{T} 
= 
 ( M_{2} A_{2}^{2} + M_{1}  B_{2}^{2} ) 
\begin{pmatrix}
c_{1}^{2}& 0 & c_{1} c_{3} \\
0 & 1/ M_{1} M_{2} & 0 \\
c_{1} c_{3} & 0 & c_{3}^{2} \\
\end{pmatrix} \, . 
\label{YTM2}
\end{align}
The mass singular values of NH and IH are 
\begin{align}
{\rm NH} &: m_{1} = 0 \, , ~~ 
m_{2} = \abs{{A_{2}^{2} \over M_{1}} + {B_{2}^{2} \over M_{2}} } \, , ~~ m_{3}  =  \abs{ M_{2} A_{2}^{2}+ M_{1}  B_{2}^{2} } (|c_{1}|^{2} + |c_{3}|^{2})  \, , \label{40} \\
{\rm IH} &: m_{1} = \abs{  M_{2} A_{2}^{2} + M_{1}  B_{2}^{2} } (|c_{1}|^{2} + |c_{3}|^{2}) \, , ~~
m_{2} = \abs{{A_{2}^{2} \over M_{1}} + {B_{2}^{2} \over M_{2}} } \, , ~~ m_{3} = 0 \, .\label{41}
\end{align}
The ratios of the singular values are 
\begin{align}
 {m_{3} \over m_{2}} &= M_{1} M_{2} (|c_{1}|^{2} + |c_{3}|^{2}) \, , 
~~ |c_{1}|^{2} + |c_{3}|^{2} = {m_{3} \over m_{2} \det M_{R}}
  \, , \\
{m_{2} \over m_{1}} &= {1 \over M_{1} M_{2} (|c_{1}|^{2} + |c_{3}|^{2}) } \, , 
~~  |c_{1}|^{2} + |c_{3}|^{2} = {m_{1} \over  m_{2} \det M_{R}} \, .
\end{align}
By using  phase transformation of the overall factor and the massless direction, 
the unitary matrix diagonalizing the mass matrix~(\ref{YTM2}) is represented as 
\begin{align}
U_{13}^{\rm NH} P & = 
\begin{pmatrix}
c_{\th} & 0 & s_{\th} e^{- i \phi} \\
0 & e^{i\a/2} & 0 \\
- s_{\th} e^{i \phi} & 0 & c_{\th}
\end{pmatrix} 
=
\begin{pmatrix}
 {|c_{3}| \over \sqrt{|c_{1}|^{2} + |c_{3}|^{2}} } & 0 & {|c_{1}| e^{i (\arg c_{1} - \arg c_{3})}\over \sqrt{|c_{1}|^{2} + |c_{3}|^{2}} }\\
0  & e^{- i \arg c_{3}} & 0 \\
 {- |c_{1}| e^{i (\arg c_{3} - \arg c_{1})}\over \sqrt{|c_{1}|^{2} + |c_{3}|^{2}} } & 0 & {|c_{3}| \over \sqrt{|c_{1}|^{2} + |c_{3}|^{2}}  }\\
\end{pmatrix}  
\, , 
\\
U_{13}^{\rm IH} P & = 
\begin{pmatrix}
c_{\th} & 0 & s_{\th} e^{- i \phi} \\
0 & e^{i\a/2} & 0 \\
- s_{\th} e^{i \phi} & 0 & c_{\th}
\end{pmatrix} 
=
\begin{pmatrix}
 {|c_{1}| \over \sqrt{|c_{1}|^{2} + |c_{3}|^{2}} } & 0 & - {|c_{3}| e^{i (\arg c_{1} - \arg c_{3})}\over \sqrt{|c_{1}|^{2} + |c_{3}|^{2}} }\\
0  & e^{-i \arg c_{1}} & 0 \\
 {|c_{3}| e^{i (\arg c_{3} - \arg c_{1})}\over \sqrt{|c_{1}|^{2} + |c_{3}|^{2}} } & 0 & {|c_{1}| \over \sqrt{|c_{1}|^{2} + |c_{3}|^{2}}  }\\
\end{pmatrix} . 
\end{align}
From this, the phenomenological parameters are found to be
\begin{align}
{\rm NH} &: \tan \th^{\rm TM_{2}} = \abs{c_{1} \over c_{3}} \, ,
~~ \phi = \arg {c_{3} \over c_{1}} \, ,  ~~ \a = - 2 \arg c_{3} \, , 
\\
{\rm IH} &: \tan \th^{\rm TM_{2}} = - \abs{c_{3} \over c_{1}} \, , 
~~ \phi = \arg {c_{3} \over c_{1}} \, ,  ~~ \a = - 2 \arg c_{1} \, . 
\end{align}
Note that again $s_{13}$ and the CP phases are determined only by certain combinations of $c_{i}$.

Substituting these parameters to the $Y_{2}$, we obtain 
\begin{align}
Y_{2}^{\rm NH} &= 
\begin{pmatrix}
0 & 0 \\
A_{2} & B_{2} \\
0 & 0 \\
\end{pmatrix} 
+  e^{- i \a/2} \sqrt{m_{3} \over m_{2} \det M_{R}}
\begin{pmatrix}
- s_{\th}^{\rm TM_{2}} e^{-i \phi} B_{2} & s_{\th}^{\rm TM_{2}} e^{-i\phi} A_{2} \\
0 & 0 \\
- c_{\th}^{\rm TM_{2}} B_{2} & c_{\th}^{\rm TM_{2}} A_{2} \\
\end{pmatrix}
 M_{R} \, ,  \\
Y_{2}^{\rm IH} &= 
\begin{pmatrix}
0 & 0 \\
A_{2} & B_{2} \\
0 & 0 \\
\end{pmatrix} 
+  e^{-i\a /2} \sqrt{m_{1} \over m_{2} \det M_{R}}
\begin{pmatrix}
- c_{\th}^{\rm TM_{2}} B_{2} & c_{\th}^{\rm TM_{2}} A_{2} \\
0 & 0 \\
s_{\th}^{\rm TM_{2}} e^{i \phi} B_{2} & -s_{\th}^{\rm TM_{2}} e^{i\phi} A_{2} \\
\end{pmatrix} 
M_{R} \, . 
\end{align}
Finally, by substituting $s_{\th}^{\rm TM_{2}} e^{i\phi} \simeq \sqrt{3/2} \, s_{13} e^{i \d}$, 
it is displayed by phenomenological parameters.
From the constraint on the mass singular values $m_{i}$~(\ref{40}) and (\ref{41}), 
there is only one free complex parameter.

\subsection{Implications for the field theoretical realization}

For the chiral solution of TM$_{1}$, $Y$ has the following form; 
\begin{align}
Y_{0}
= 
\begin{pmatrix}
0 & 0 \\
A_{2} & B_{2} \\
A_{3} & B_{3}
\end{pmatrix} , 
~~~ 
Y =
\begin{pmatrix}
a_{2} & b_{2} \\
a_{2} + a_{3} & b_{2} + b_{3} \\
a_{2} - a_{3} & b_{2} - b_{3}
\end{pmatrix} \, .
\end{align}
The texture is realized by vevs of flavons $\vev{\phi_{2}} \propto (1,1,1)$ and $\vev{\phi_{3}} \propto (0,1,-1)$ that have the $+1$ charge under the $Z_{2}$ symmetry of $S_{1}$ (see Eq.~(\ref{diagS12})). 
Thus, such a model is possible by prohibiting the flavon $\vev{\phi_{1}} \propto (-2,1,1)$ with the charge $-1$.

On the other hand, since the nontrivial solutions have intricate conditions between $Y$ and $M_{R}$, 
it is generally difficult to realize these textures.  
However, there are approximate solutions with a simple construction.
For the Yukawa matrix to be in the above $Z_{2}$ eigenstates, 
$ A_{1} = 0$ or $B_{1} = 0$ is required. Then the symmetry condition~(\ref{condTM1}) is reduced to 
\begin{align}
{A_{2} \over B_{2}} = {A_{3} \over B_{3}} = {-M_{11} B_{1} + M_{12} A_{1} \over - M_{12} B_{1} + M_{22} A_{1}}  
= { M_{11}  \over M_{12}  } ~~ {\rm or} ~~ 
 { M_{12}  \over M_{22} } \, . 
\end{align}
The small lightest mass $M_{1} \ll M_{2}$ acompanies an approximate chiral symmetry 
 of $M_{R}$ that ensures  $|M_{11}|  \ll |M_{12}|$ and/or $|M_{12}|  \ll |M_{22}|$ in some basis.  
If $A_{1}$ or $B_{1}$ are small enough in this basis, each column vectors of $Y''$ is almost $Z_{2}$ eigenstates and this texture will be realized by the above vevs $\vev{\phi_{i}}$.
%
In this case the breaking of $Z_{2}$ symmetry is about $O(M_{1(1,2)} / M_{(1,2)2})$ and 
the accuracy of TM$_{1,2}$ mixing is good enough if $M_{R}$ is sufficiently hierarchical.

\section{Summary}

In this paper, we concisely represent the general solution for the TM$_{1,2}$ mixing in the type-I minimal seesaw mechanism    and show phenomenological constraints of the Yukawa matrix of neutrinos $Y$. 
First, conditions of $Z_{2}$ symmetry associated with TM$_{1,2}$ restrict two complex parameters.  
In addition, three complex parameters are constrained from five phenomenological parameters (two neutrino masses $m_{i}$, the mixing angle $\theta_{13}$, the Dirac phase $\delta$, and the Majorana phase $\alpha$) and the overall phase.
As a result, there is only one free complex parameter for $Y$, except for the right-handed neutrino masses $M_{1,2}$.

For the TM$_{1}$ mixing with normal hierarchy, the symmetry condition is imposed only on $Y$, and the physical parameters depend on $M_{R}$. In other situations, the dimensionless parameters $(\theta_{13}, \delta$ and $\alpha)$ do not depend on $M_{R}$ but only on certain combinations of the components of $Y$.

This property is because the two massive neutrinos with NH belong to the same representation of $Z_{2}$ symmetry of TM$_{1}$  and its eigenspace is two-dimensional. 
In other situations, the two massive neutrinos have different charges $\pm1$ of the $Z_{2}$ symmetry, 
and the diagonalization is solved easily for the one-dimensional eigenspaces.



\bibliographystyle{bib/h-physrev50}


\end{document}